\pdfoutput=1
\documentclass[a4paper,11pt]{article}
\usepackage{jinstpub} % for details on the use of the package, please
\title{Radiation hard pixel sensors using high-resistive wafers in a 150\,nm CMOS processing line}
\author[a,1]{D.-L.~Pohl\note{Corresponding author.},}
\author[a]{T.~Hemperek,}
\author[a]{I.~Caicedo,}
\author[b]{L.~Gonella,}
\author[a]{F.~H{\"u}gging,}
\author[a]{J.~Janssen,}
\author[a]{H.~Kr{\"u}ger,}
\author[c]{A.~Macchiolo,}
\author[a]{N.~Owtscharenko,}
\author[d]{L.~Vigani,}
\author[a]{and~N.~Wermes}

\affiliation[a]{University of Bonn,\\
Physikalisches Institut, Nu{\ss}allee 12, D-53115 Bonn, Germany}
\affiliation[b]{University of Birmingham,\\
School of Physics and Astronomy, Edgbaston, Birmingham, B15 2TT, UK}
\affiliation[c]{Max Planck Institut for Physics,\\
Werner-Heisenberg Institut, F{\"o}hringer Ring 6, D-80805 Munich, Germany}
\affiliation[d]{University of Oxford,\\
Denys Wilkinson Building, Keble Road, Oxford, OX1 3RH, UK}

\emailAdd{pohl@physik.uni-bonn.de}

\abstract{Pixel sensors using 8$''$ CMOS processing technology have been designed and characterized offering the
benefits of industrial sensor fabrication, including large wafers, high throughput and yield, as well as low cost.
The pixel sensors are produced using a 150\,nm CMOS technology offered by LFoundry in Avezzano. The technology
provides multiple metal and polysilicon layers, as well as metal-insulator-metal capacitors that can be employed for AC-coupling and redistribution layers.
Several prototypes were fabricated and are characterized with minimum ionizing particles before and after irradiation
to fluences up to 1.1\,$\times$\,10$^{15}$~n$_{\rm eq}$\,cm$^{-2}$. The CMOS-fabricated sensors perform equally well as standard pixel
sensors in terms of noise and hit detection efficiency. AC-coupled sensors even reach 100\% hit efficiency in a 3.2\,GeV electron beam before irradiation.
}

\begin{document}
\maketitle
\flushbottom

%=================================================
\section{Introduction}\label{sec:intro}
%=================================================
Hybrid pixel detectors~\cite{pixel_book} have matured to a level that they are the main contenders for tracking and vertexing near the interaction point at experiments
of the planned LHC upgrade~(\mbox{HL-LHC},~\cite{hllhc}). The surface covered by pixel detectors will
likely increase from less than 2\,m$^2$ at present to 10--20\,m$^2$ in the future.
The detectors will be arranged in several concentric cylindrical layers, with radii between 3--30\,cm, and many square meters of disk layers in the forward and backward region of the detectors.

Industrial CMOS processes on large, high-resistive wafers enable sensor designs with high yield and high throughput at comparatively low cost. This technology is therefore of interest also for large area
pixel or strip detectors, not only for readout IC fabrication but also for the particle sensing elements, the \emph{sensors}.
Furthermore, various parts of the CMOS processing technology like multiple metal layers, polysilicon layers, as well as metal-insulator-metal (MIM) capacitors can be employed for special
sensor features that are otherwise mostly not available. Examples are AC-coupling of the sensor to the readout electronics or obtaining a significant module simplification
by implementation of redistribution layers for signals and service voltages on the sensor itself rendering a module flex much simpler or even obsolete. A redistribution of bump connections by a dedicated metal
layer can decouple the sensor pixel pitch from the readout pixel pitch thus enabling a module design without elongated inter-gap pixels.
As the CMOS sensors include no active electronic circuitry (transistors), they are termed \emph{passive CMOS sensors}.

In this paper we present radiation hard passive CMOS pixel sensors fabricated in a 150\,nm CMOS process offered by LFoundry~\cite{LFoundry_web} on high-resistive wafers (> \,2\,k$\Omega$-cm). A number of prototypes have been
characterized with respect to production quality, depletion depth, signal and noise levels, pixel capacitance, and detection efficiency before and after irradiation up to fluences of 1.1\,$\times$\,10$^{15}$~n$_{\rm eq}$\,cm$^{-2}$.

%=================================================
\section{Pixel sensor design}\label{sec:sensor_design}
%=================================================
\begin{figure}[b]
%\internallinenumbers
\centering
	\includegraphics[width=0.90\textwidth]{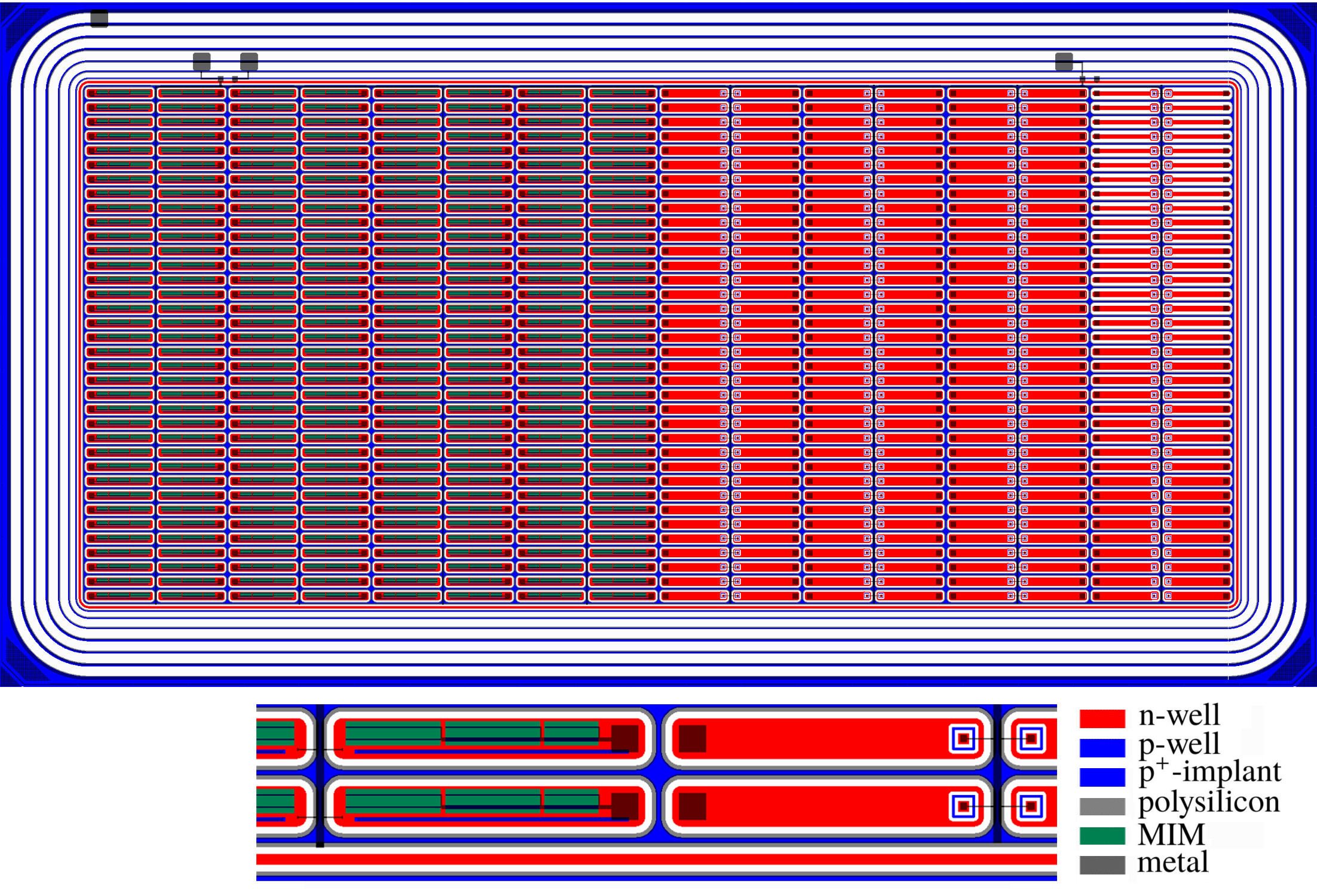}
\caption[sensor design]{Pixel sensor matrix with 576 pixels surrounded by an n-well ring and subsequently
7 p-type guard rings (top). The bottom shows 4 edge pixels with bump pads. (left) Pair of AC pixels with a large area fraction filled by MIM layers providing the capacitive coupling; 
(right) pair of DC pixels containing dedicated implants for punch through biasing (bias dots).	\label{fig:sensor_design}}
\end{figure}
Prototype n-in-p pixel sensor matrices have been produced in the LFoundry 150\,nm CMOS process on Czochralski p-type wafers having a foundry specified resistivity of typically \mbox{4--5\,k$\Omega$-cm}
(minimum \,2\,k$\Omega$-cm). The pixel matrix measures 1.8\,$\times$\,4\,mm$^2$ and is divided into 16\,$\times$\,36 pixels with a size of 50\,$\times$\,250\,$\upmu$m$^2$. The matrix is surrounded by an
n-implantation bounding the active volume of the edge pixels, followed by seven p-implantation guard rings to isolate the electrodes from the high voltage at the sensor edge (figure~\ref{fig:sensor_design}).
Half of the pixels are DC coupled with punch-through biasing and the other half is AC coupled by a 3.2\,pF MIM capacitor. AC pixels are biased via a 15\,M$\Omega$ polysilicon resistor present in each pixel.
The pixel layout follows a standard planar pixel design used, for instance, for the ATLAS IBL planar sensors~\cite{Gossling:2011zz}, with a 30\,$\upmu$m wide n-implantation as readout electrode and a 20\,$\upmu$m gap between pixels to all sides.
For the two outermost columns of the DC half, the n-implantation width has been varied in the range between 15\,$\upmu$m and 30\,$\upmu$m, to allow for
optimization studies of the fill factor with respect to input capacitance and charge-collection efficiency. All pixels are isolated by a 4\,$\upmu$m p-stop grid implemented
below field plates made of a low resistive polysilicon layer. The intention of the field plates is to flatten spikes in the electric field occurring after irradiation, hence improving the breakdown
behavior~\cite{tomasz_thesis_2017}. A high resolution mask set was used for processing as the submission was shared with other monolithic active CMOS devices (DMAPS~\cite{Kishishita:2015jda}). The use of such (expensive) mask sets is likely not needed in a dedicated passive sensor production run.
The sensors have been thinned to 100\,$\upmu$m and 300\,$\upmu$m and then backside processed providing a p-implant and a metallization layer to allow bias voltage application from the backside.
The very thin 100\,$\upmu$m wafer was processed by IBS in France~\cite{ibs_web} including a thinning step in a TAIKO process~\cite{taiko}. In this process the wafer edge is left thicker to create a self-sustainable wafer for further processing without the need of a handling wafer.
%

%=================================================
\section{Device assembly and characterization procedures}\label{sec:setup}
%=================================================
The sensor was bonded to the ATLAS FE-I4 pixel readout-chip~\cite{FE-I4} via fine pitch solder bump bonding at Fraunhofer IZM in Berlin~\cite{IZM_Klein_2008}. The FE-I4 chip was tuned to threshold values similar to those 
used for the IBL layer of the ATLAS pixel detector corresponding to approximately 3000\,e$^-$. The chip was readout with the USBpix~2.0 hardware~\cite{Backhaus:2011svd} in combination with \emph{pyBAR} software,
a Python based readout- and test-system for the ATLAS FE-I4~\cite{pybar}. The FPGA firmware modules and interfaces were provided by \emph{basil}, a modular data-acquisition and system-testing framework~\cite{basil}.

The sensor efficiencies were determined in several test-beam campaigns using minimum ionizing particles. Electrons were used with energies between 2.5\,GeV and 3.2\,GeV at the electron stretcher accelerator (ELSA) in
Bonn~\cite{Heurich:2016ilc} and 120 GeV pions at the super proton synchrotron (SPS) at CERN~\cite{Synchrotron:1997188}. A fast particle telescope based on two ATLAS FE-I4 modules~\cite{IBL_modules_2012} was used
for high-rate test-beams at ELSA to determine the detection and charge collection efficiencies for different bias and threshold settings with high statistics. A high resolution telescope (EUDET~\cite{EUDET_Telescope})
based on six Mimosa26 planes~\cite{HuGuo:2010zz} and one FE-I4 hybrid pixel time-reference plane was used for the high-energy test-beam setup at CERN to
map with sub-pixel resolution. The synchronization between data from FE-I4 telescope planes and Mimosa26 planes was established by means of the JUDITH software~\cite{Judith:2014baa}. This synchronization is difficult
and error prone, since the event integration time of the Mimosa planes is much longer than the integration time of the ATLAS FE-I4 (115\,$\upmu$s vs. 0.4\,$\upmu$s). A mismatch between tracks recorded in the Mimosa telescope with
hits in the FE-I4 time reference plane assigns a wrong time stamp to these tracks during reconstruction. Every mismatched track artificially decreases the determined absolute efficiency. Consequentially, the absolute
efficiency quoted in this work was always determined with the FE-I4 based telescope, where synchronization issues due different event integration times do not exist.

The subsequent steps of a test-beam analysis involving hit clustering, track finding and fitting were carried out using \emph{testbeam analysis}~\cite{testbeam_analysis}, a novel python package.
The performance of \emph{testbeam analysis} was verified by simulation as well as by analyzing test-beam data from well understood and formerly characterized ATLAS IBL pixel modules~\cite{IBL_modules_2012}.

To determine the charge spectra of the sensor a new charge reconstruction method was developed (\emph{TDC method}). This method overcomes the limitation of the front-end to measure charge
spectra with sufficient resolution by sampling the charge signal with a faster external clock utilizing the FPGA of the readout system~\cite{TDC-method}.

One sensor has been irradiated with 24\,MeV protons at the Birmingham irradiation facility~\cite{Dervan:2015pua} in steps to 0.18\,$\times$\,10$^{15}$~n$_{\rm eq}$\,cm$^{-2}$ and 1.14\,$\times$\,10$^{15}$~n$_{\rm eq}$\,cm$^{-2}$. After each step the sensor
was annealed for 80 minutes at 60$^\circ$C. The total fluence was measured with about 10\% uncertainty by post-irradiation dosimetry on nickel foils attached to the device during irradiation~\cite{nickel_foil}.

%=================================================
\section{Results}\label{sec:results}
%=================================================
\subsection{I-V curves}\label{sec:IV-curves}
\begin{figure}[h]
%\internallinenumbers
	\centering
		\includegraphics[width=0.47\textwidth]{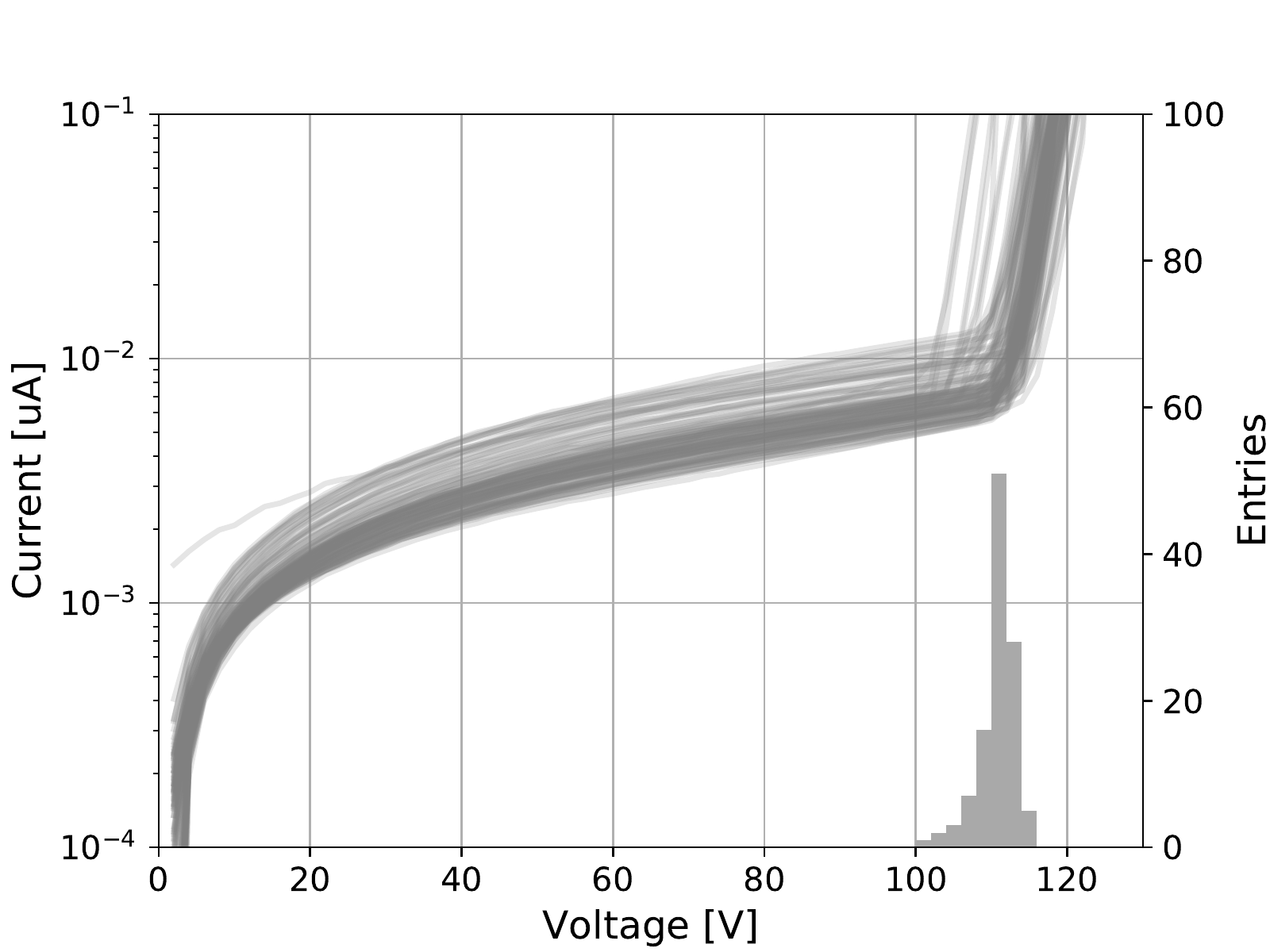}
	\qquad
		\includegraphics[width=0.47\textwidth]{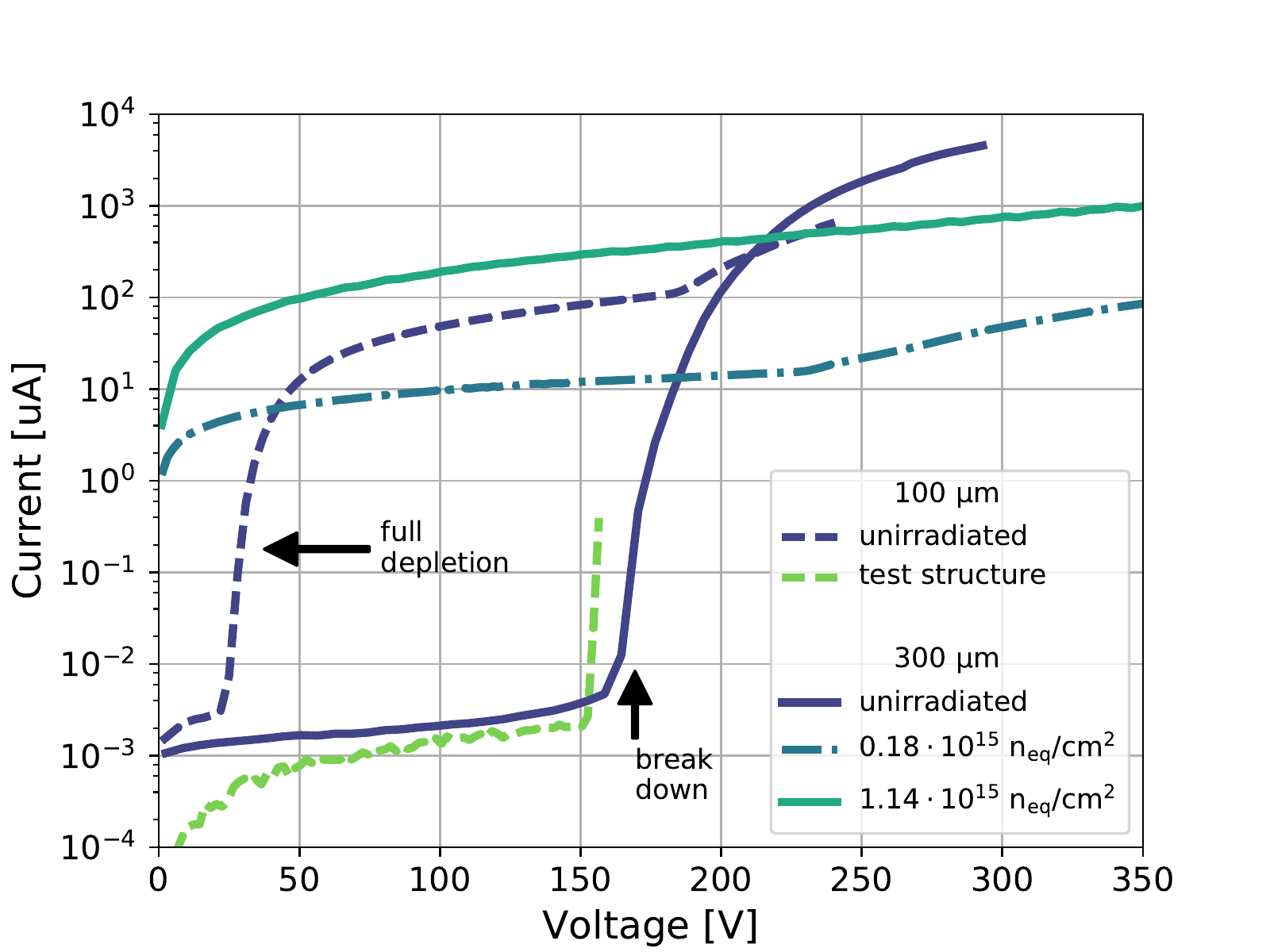}
\caption{I-V measurements of LFoundry CMOS sensors: (left) I-V curves before irradiation and dicing of 114 sensors located on the same wafer thinned to 300\,$\upmu$m. One sensor suffering a short circuit due to a processing
error has been excluded from the plot. (right) I-V curves of two pixel module assemblies. I-V measurements for two different thicknesses (100\,$\upmu$m (dark dashed line) and 300\,$\upmu$m (dark solid line) and
two different irradiation levels (0.18\,$\times$\,10$^{15}$~n$_{\rm eq}$\,cm$^{-2}$ (dash-dotted line) and 1.14\,$\times$\,10$^{15}$~n$_{\rm eq}$\,cm$^{-2}$ (light solid line)) are shown. The light dashed line shows an unbonded test structure
treated with plasma etching after backside grinding to mitigate the current increase at full depletion. The results are normalized to 20$^\circ$C temperature.\label{fig:IV-curves}}
\end{figure}

To examine the process- and design quality I-V curves (current versus reverse-bias voltage) of 114 sensors have been obtained. The sensors were located on one wafer half of a 300\,$\upmu$m thin wafer and were contacted at
the bias grid by needles on a probe station. The curves in figure~\ref{fig:IV-curves} (left) show very similar electrical characteristics for all sensors. The breakdown requires bias voltages in excess
of 100\,V for all 114 sensors and is independent of the sensor's position on the wafer. Only one sensor showed a short due to a processing error. The location of the breakdown was determined via emission microscopy on 5
sensors of the same wafer after dicing and it occurs at the bias dots of the DC pixels.
The right side of figure~\ref{fig:IV-curves} depicts results of the IV measurements of two fully assembled pixel modules after thinning, backside processing, and solder bonding to the ATLAS FE-I4 chip.
Backside processing after thinning included a p$^+$ implantation plus metallization, but was done without an etching step after backside grinding.
It has been observed before~\cite{Okihara:2015rns} that excessive leakage current can be mitigated by applying etching after grinding. The absence of this treatment is likely the cause for the observed
behavior at 24\,V for the 100\,$\upmu$m device (figure~\ref{fig:IV-curves}, arrow \emph{full depletion}). At this voltage the depletion region reaches the backside, in agreement with calculations assuming a resistivity
of about 5\,k$\Omega$-cm (see section~\ref{sec:depletion}). In order to prove this hypothesis a dedicated test structure with the same pixel pitch, guard rings, and implantations was designed and thinned to 100\,$\upmu$m and the
backside was treated with plasma etching. It can be seen in figure~\ref{fig:IV-curves} (\emph{test structure}) that the additional current increase is not occurring anymore when the depletion zone touches the backside.
The additional current increase at about 160--180\,V for the unirradiated 100\,$\upmu$m and 300\,$\upmu$m is likely a breakdown at the bias dots of the DC pixels.
After irradiation of the sensor assembly the breakdown voltage increases. At an irradiation level of 1.14\,$\times$\,10$^{15}~$n$_{\rm eq}$\,cm$^{-2}$ voltages in excess of 700\,V have been applied without observing
any breakdown.
\subsection{Signal and noise}\label{sec:SN}
\begin{figure}[h]
%\internallinenumbers
	\centering
		\includegraphics[width=0.47\textwidth]{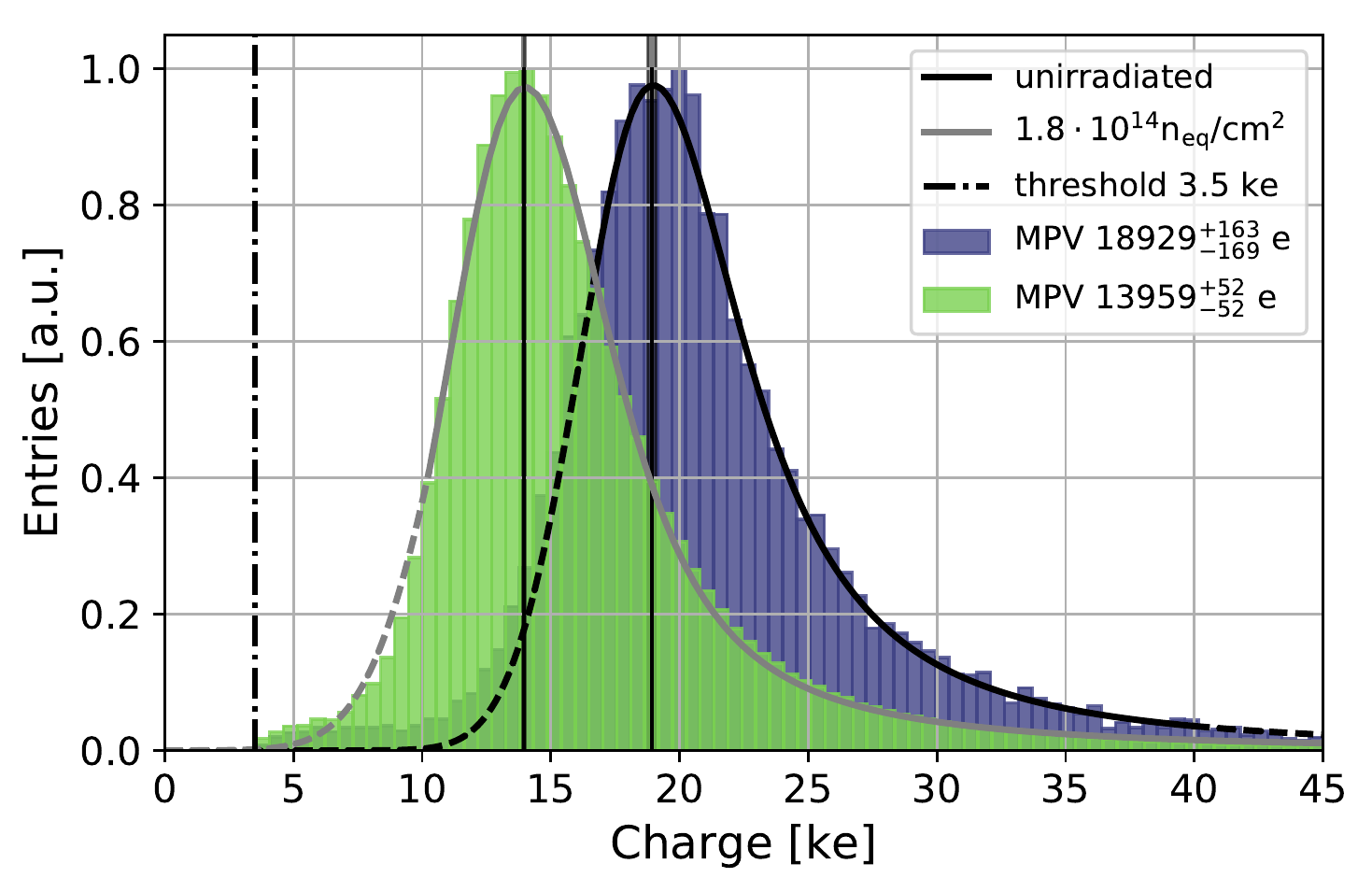}
	\qquad
		\includegraphics[width=0.47\textwidth]{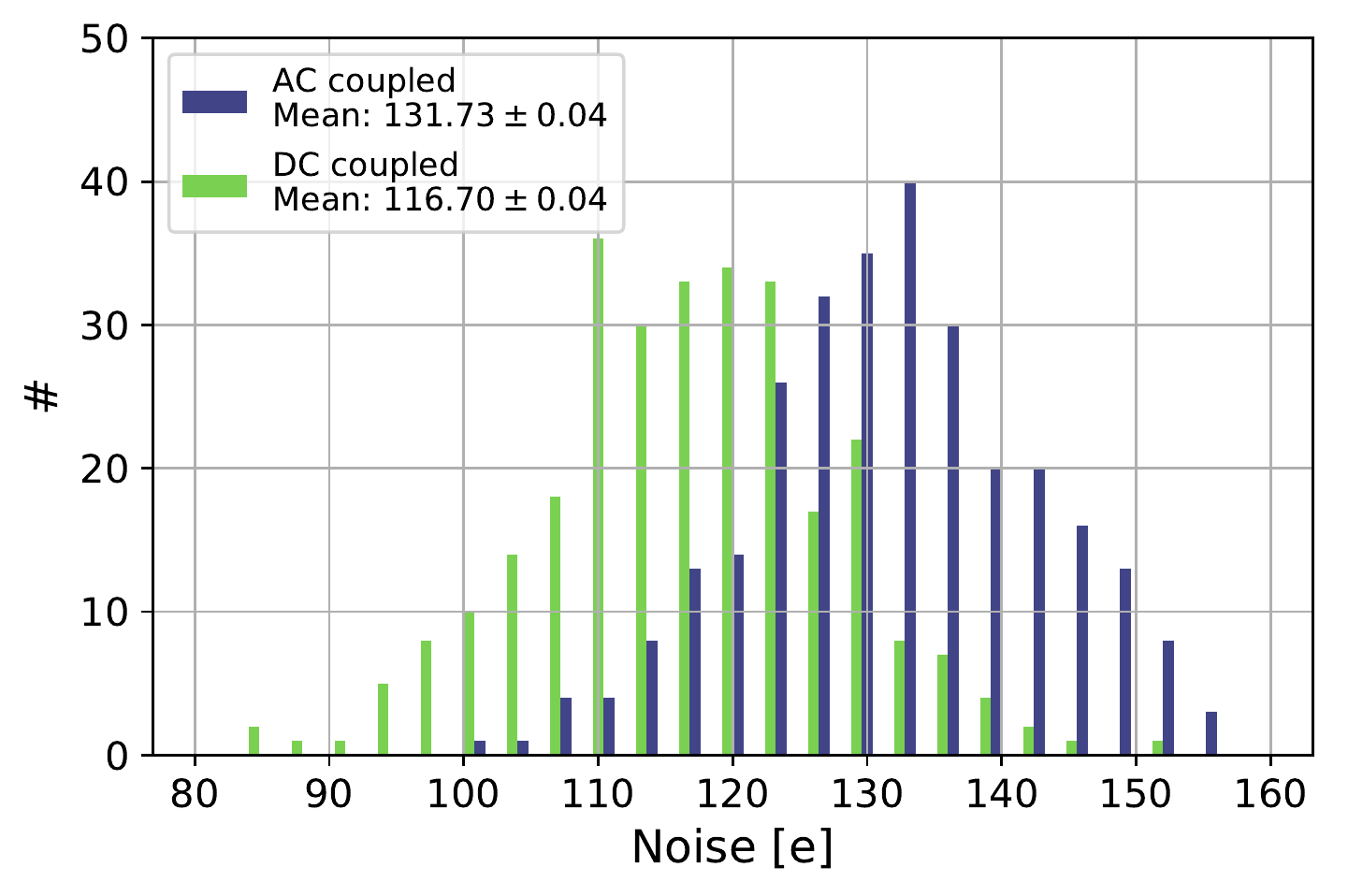}
\caption{Signal and noise for a 300\,$\upmu$m thick LFoundry passive pixel sensor at 160\,V bias bump bonded to the ATLAS FE-I4. (left) Measured charge distributions (normalized) of single hit clusters of 3.2~GeV
electrons before irradiation and after 0.18\,$\times$\,10$^{15}$~n$_{\rm eq}$\,cm$^{-2}$. The charge was measured using the TDC method (see text) and fitted with a Landau-Gauss convolution
to determine its MPV. (right) The equivalent noise charge distributions for AC (dark) and DC (light) coupled pixels. The mean noise for these couplings is stated in the legend. \label{fig:signal_and_noise}}
\end{figure}

Figure~\ref{fig:signal_and_noise} (left) shows single pixel charge distributions of the 300\,$\upmu$m thick sensor obtained in test-beams (3.2\,GeV electrons) before irradiation at 160\,V bias and after
0.18\,$\times$\,10$^{15}$~n$_{\rm eq}$\,m$^{-2}$ at 250\,V. A similar measurement at the highest fluence of 1.14\,$\times$\,10$^{15}$~n$_{\rm eq}$\,cm$^{-2}$ suffered an electric discharge disabling the FE-I4 readout chip functionality, and is thus not available.

Figure~\ref{fig:signal_and_noise} (right) shows distributions of the noise for AC and DC coupled pixels determined using the internal charge injection circuitry of the ATLAS FE-I4 when a 300\,$\upmu$m sensor at
160\,V bias is attached. The distributions have mean values of ENC~=~117\,$e^-$~(DC) and 132\,$e^-$~(AC).
This should be compared to the noise performance of planar IBL sensors (ENC\,$\approx$\,120\,$e^-$~\cite{IBL_modules_2012}, C$_D$~=~117\,fF~\cite{pixcap2013}) and IBL 3D-Si sensors
(ENC\,$\approx$\,140\,$e^-$~\cite{IBL_modules_2012}, C$_D$~=~180\,fF~\cite{pixcap2013}) showing that the noise performance of the passive CMOS sensor investigated here is comparable. Hence we conclude that the DC coupled
pixels have an input capacitance less than 120\,fF and the AC coupled pixel less than 180\,fF. The larger value for the AC coupled sensors arises from the parasitic capacitance of the polysilicon layer
by which the bias resistor is implemented. The values for the bias resistor and coupling capacitance were not optimized in the design, but chosen as large as possible, leaving room for input
capacitance optimization in future designs.

\subsection{Depletion depth}\label{sec:depletion}
Assuming that in silicon a minimum ionizing particle creates about 71\,e/h pairs per micron (most probable value, from GEANT4 simulation) and that only the depleted volume of the sensor contributes to the charge
signal (due to the fast response time of the FE-I4) we determine the depletion depth and thus the p-type bulk resistivity of the passive CMOS sensors using
\begin{equation}
d = \sqrt{\frac{2\epsilon_0 \epsilon_r}{eN_{\rm eff}} \,
\left( V_{\mathrm{bi}} + V_{\mathrm{bias}} \right) }\qquad \Rightarrow \qquad d\left[\upmu{\rm m}\right] \approx 0.3\, \sqrt{\rho\left[\Omega {\,\rm cm} \right] \cdot V_{\mathrm{bias}}\left[\mathrm{V}\right] }
\label{eq:depletion_depth}
\end{equation}
where $d$ is the depletion depth, $N_{\rm eff}$ the effective doping concentration, $\epsilon_0$ the vacuum permittivity and $\epsilon_r$ the relative permittivity of silicon.
\begin{figure}[ht]
%\internallinenumbers
\centering
	\includegraphics[width=0.9\textwidth]{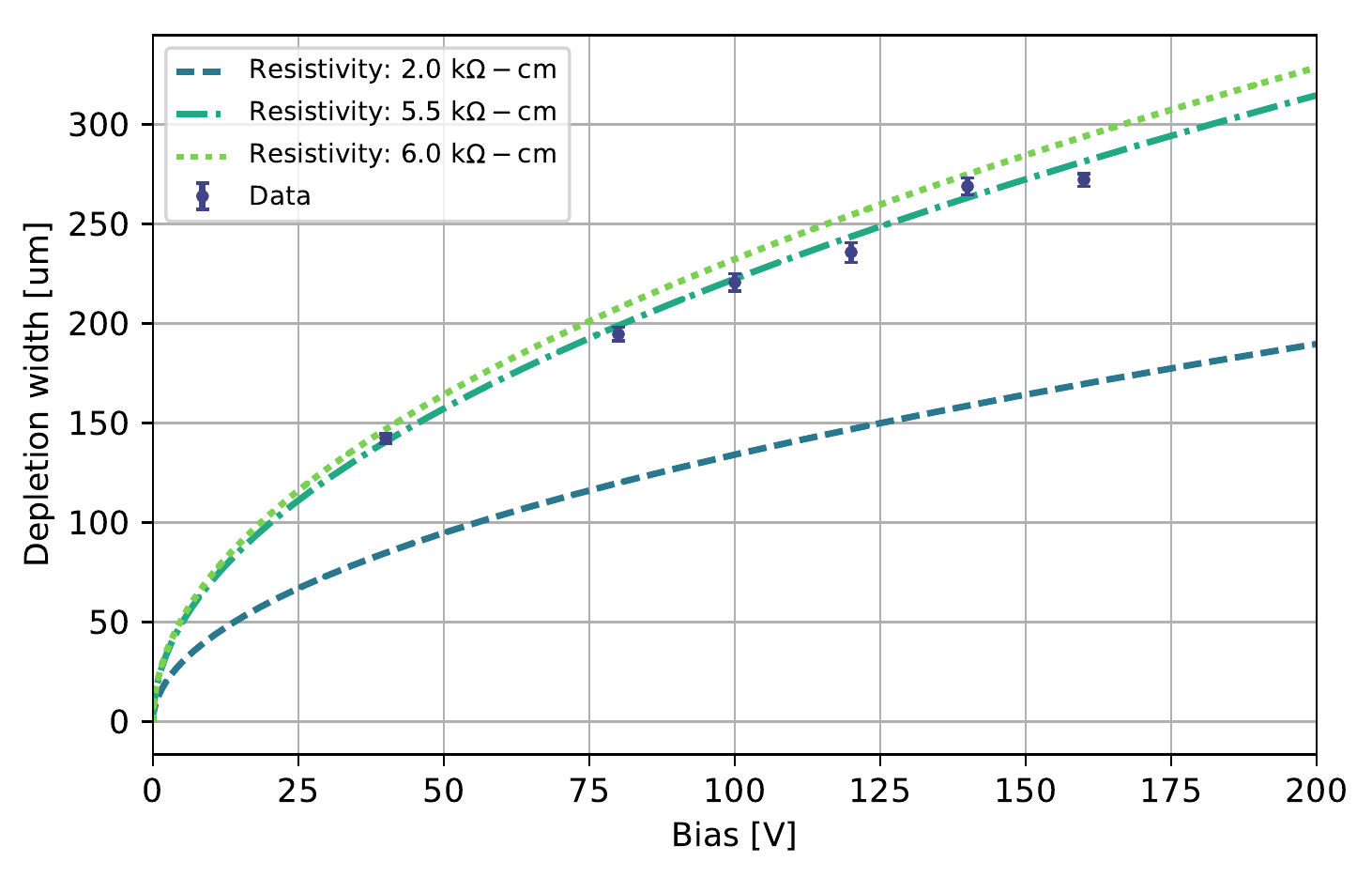}
\caption[depletion_depth]{Depletion depth as a function of bias voltage for the 300\,$\upmu$m thick LFoundry passive pixel sensor. The depletion depth was calculated using Langau fits as
in figure~\ref{fig:signal_and_noise} assuming 71\,e/h pairs created per micrometer. Theoretical depletion curves for three bulk resistivities given by \ref{eq:depletion_depth} are shown for comparison. \label{fig:depletion_depth}}
\end{figure}

Figure~\ref{fig:depletion_depth} shows the measured depletion depth assuming 71\,e/h pairs created per micron
using the most probable value of measured charge distributions as in figure~\ref{fig:signal_and_noise} (left) extracted by a fit to a so-called \emph{Langau} distribution~\cite{Langau}, a Landau distribution~\cite{Kolbig:1983uf} folded with a Gaussian.
The measured depletion depths at different bias voltages are compared to theoretical curves obtained from \eqref{eq:depletion_depth} assuming three different values for the (p-type) bulk resistivity.
The error bars of the measurements are defined by the fit error and the uncertainty of the calibration of the charge-injection circuitry of the ATLAS FE-I4. One can conclude that the resistivity of the bulk material
of the measured wafer is at least 4--5\,k$\Omega$-cm, a value compatible with the range 
%(min. 2\,k$\Omega$-cm, typical 4--5\,k$\Omega$-cm) 
quoted by the vendor.

\subsection{Hit detection efficiency}
\begin{figure}[t]
%\internallinenumbers
\begin{center}
	\includegraphics[width=0.9\textwidth]{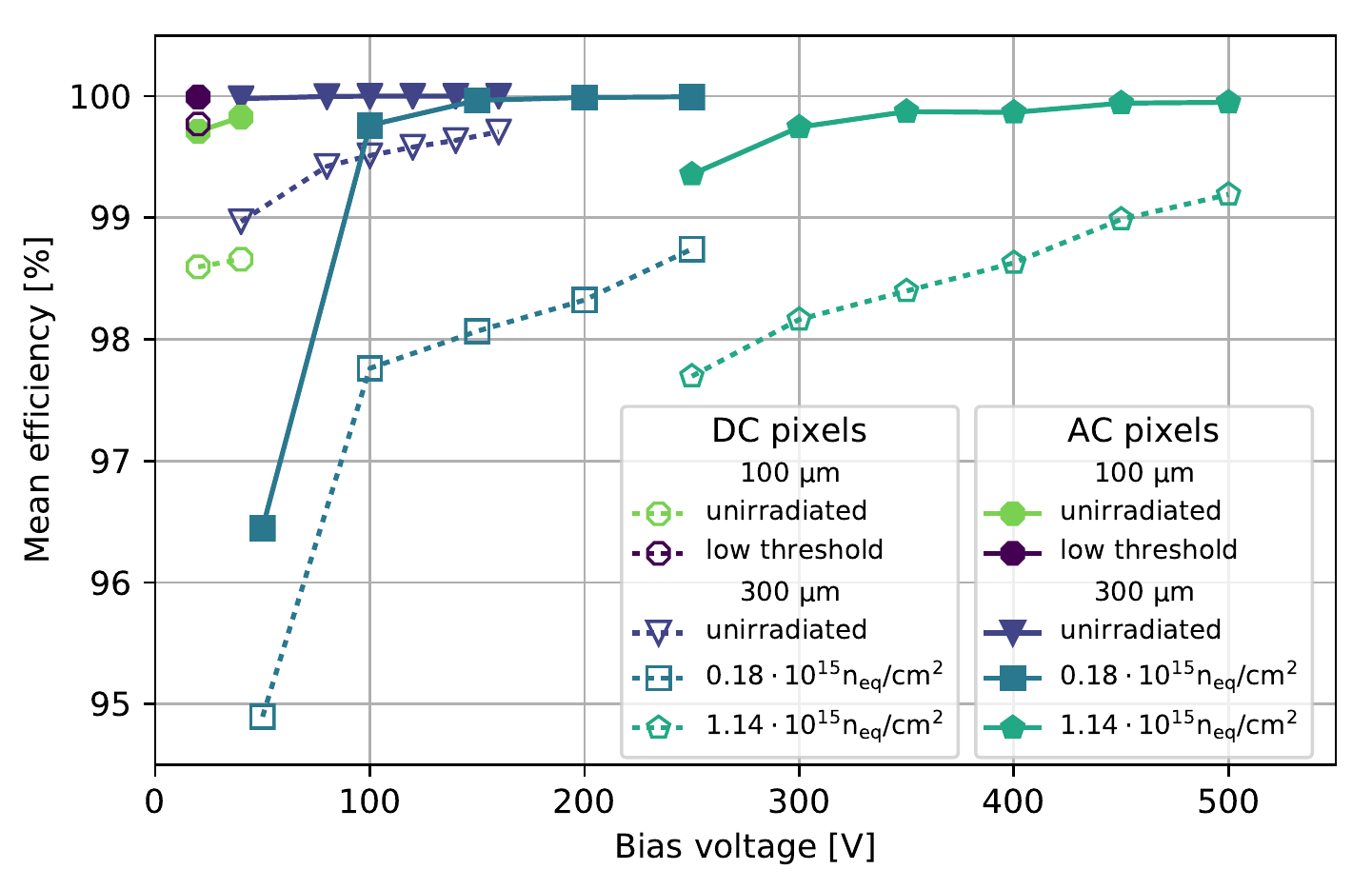}
\caption[efficiency]{Hit detection efficiency of 100/300\,$\upmu$m thick pixel sensors for different bias voltages and different levels of radiation damage. Each point represents the mean
efficiency of at least 50 center pixels. The statistical error bars are too small to visualize and are thus not shown.	\label{fig:efficiency}}
\end{center}
\end{figure}
The hit detection efficiency has been determined using 3.2\,GeV electrons for sensors thinned to
100\,$\upmu$m (unirradiated) and 300\,$\upmu$m (three levels of radiation: 0, 0.18\,$\times$\,10$^{15}$~n$_{\rm eq}$\,cm$^{-2}$ and 1.14\,$\times$\,10$^{15}$~n$_{\rm eq}$\,cm$^{-2}$).
The results are shown in figure~\ref{fig:efficiency}. The hit efficiency reaches values of $98.7\%$/$99.8\%$ for the DC/AC coupled 100\,$\upmu$m sensor pixels at a standard threshold of about
3000~electrons and increases to $99.8\%$/$>99.9\%$ after reducing the threshold to about 1500~electrons.
After irradiation the efficiency reaches values $>99.9\%$ for AC coupled sensors at bias voltages above 350\,V. The DC coupled sensor efficiency is somewhat lower ($>99\%$ at 450\,V) due to
the \textquotesingle blind\textquotesingle\ area taken by the punch-through biasing dot.
\begin{figure}[b]
%\internallinenumbers
\centering
	\includegraphics[width=1.\textwidth]{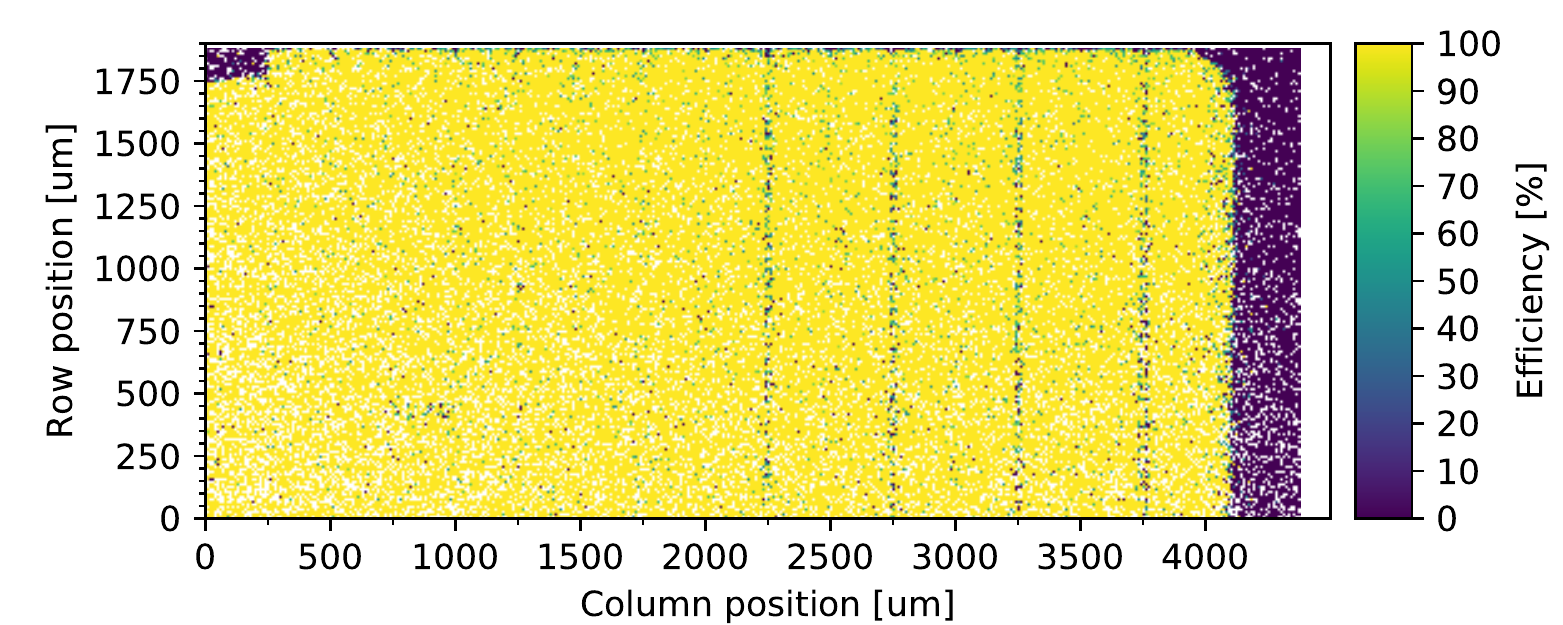}
\caption[Efficiency_map]{Hit detection efficiency map of the LFoundry passive pixel sensor with 300\,$\upmu$m  thickness at 160\,V bias. Bins of 10\,$\upmu$m x 10\,$\upmu$m size have
been used and are labeled by a white color where the efficiency determination is not possible due to low statistics. The efficiency is not correct on an absolute scale (see section~\ref{sec:setup}), but
shows the efficiency loss at the location of the bias dots. The upper left inefficient area are two disabled pixels during analysis and the inefficient area to the right marks the edge of the active sensor
area.\label{fig:efficiency_map}}
\end{figure}
This effect can be seen in figure~\ref{fig:efficiency_map}, showing the efficiency map over all pixels determined with 120\,GeV pions. The map depicts a histogram with 10\,$\upmu$m~x~10\,$\upmu$m bins of the intersection points of reconstructed particles tracks with the sensor. The ratio of tracks that created a hit in the sensor, within an association distance of 500\,$\upmu$m, divided by the total number
of tracks gives the efficiency per bin. While the efficiency of the AC coupled pixels is homogenous, it is possible to observe an efficiency loss at the bias dots located in the DC coupled pixels.
The absolute efficiency is underestimated (see section~\ref{sec:setup}) explaining a lower efficiency in comparison to figure~\ref{fig:efficiency}.

% =================
\section{Conclusions}\label{sec:conclusions}
% =================
In this paper we presented pixel sensors fabricated in a CMOS process line using high ohmic ($\sim$5k$\Omega$-cm) substrate wafers. Employing CMOS technology provides several benefits for large area pixel detectors
such as those planned for the HL-LHC upgrade at CERN, namely: low cost, large through-put production on 8$''$ or even 12$''$ wafers, and design flexibility (e.g. AC coupling, routing) while keeping the charge collection
properties of standard sensors. The wafers were fabricated using LFoundry 150\,nm technology. The sensors were thinned to 300\,$\upmu$m and
100\,$\upmu$m thickness and were backside processed. A sensor breakdown after irradiation to more than 10$^{15}$~n$_{\rm eq}$\,cm$^{-2}$ did not occur up to a maximum applied voltage of 700\,V.
Measured signal and noise performance was found to be very similar to those of planar pixel sensors currently used in the ATLAS experiment.
The hit-efficiency after irradiation was measured in an electron test-beam to be above 99.9\% (99.1\%) for AC- (DC)-coupled sensor-chip assemblies.

\acknowledgments
We would like to thank Julie Segal for the emission microscopy to localize the sensor breakdown. We also thank Ion Beam Services (IBS) in France and LFoundry in Italy for their kind support and useful feedback during sensor
production and backside processing.
This work was supported by the German Ministerium f{\"u}r Bildung, Wissenschaft, Forschung und Technologie (BMBF) under contract no. 05H15PDCA9,
by the European Union's Horizon 2020 project AIDA-2020 under grant agreement no. 654168,
and a Marie Sk\l{}odowska-Curie ITN Fellowship of the European Union's Horizon 2020 program under grant agreement no. 675587-STREAM.

\end{document}